\documentclass[aps,pra,preprint,superscriptaddress,longbibliography]{revtex4-1}

\usepackage{amsmath}
\usepackage{amsthm}
\usepackage{amsfonts}
\usepackage{amssymb}
\usepackage{xcolor,graphicx}
\usepackage{tikz}
\usepackage{color}
\usepackage[pdftex]{hyperref} %para ligar las referencias
\usepackage{longtable}
\usepackage{array}
\usepackage{dsfont}

\begin{document}
	
%--------------------------------------------------------------------------------
	\title{Symmetric supermodes in cyclic multicore fibers}
	
	\author{Benjam\'in Jaramillo \'Avila}
	\email[e-mail: ]{jaramillo@inaoep.mx}
	\affiliation{CONACYT-Instituto Nacional de Astrof\'{i}sica, \'{O}ptica y Electr\'{o}nica, Calle Luis Enrique Erro No. 1. Sta. Ma. Tonantzintla, Pue. C.P. 72840, M\'{e}xico.}

	\author{Javier Naya Hern\'andez}
	\affiliation{Tecnologico de Monterrey, Escuela de Ingenier\'ia y Ciencias, Ave. Eugenio Garza Sada 2501, Monterrey, N.L., M\'exico, 64849.}

	\author{Sara Mar\'ia Toxqui Rodr\'iguez}
	\affiliation{Academia Mexicana de Ciencias, Verano de la Investigaci\'on Cient\'ifica - Tecnologico de Monterrey, Escuela de Ingenier\'ia y Ciencias, Ave. Eugenio Garza Sada 2501, Monterrey, N.L., M\'exico, 64849.}
	
	\author{Blas Manuel Rodr\'iguez-Lara}
	\affiliation{Tecnologico de Monterrey, Escuela de Ingenier\'ia y Ciencias, Ave. Eugenio Garza Sada 2501, Monterrey, N.L., M\'exico, 64849.}
	\affiliation{Instituto Nacional de Astrof\'{i}sica, \'{O}ptica y Electr\'{o}nica, Calle Luis Enrique Erro No. 1. Sta. Ma. Tonantzintla, Pue. C.P. 72840, M\'{e}xico.}
	
	\date{\today}
	
	\begin{abstract}
		Nearest-neighbor coupled-mode theory is a powerful framework to describe electromagnetic-wave propagation in multicore fibers but it lacks precision as the separation between cores decreases.	
		We use abstract symmetries to study a ring of evenly distributed identical cores around a central core, 
		a common configuration used in telecommunications and sensing. 
		We find its normal modes and their effective propagation constants while including the effect of all high-order inter-core couplings.
		Finite-element simulations support our results to good agreement. 
		Only two of these effective modes involve fields in all the cores.
		These two modes display opposite-sign phase configurations between the fields in the external cores and the central core. 
		These two modes still appear in the limit where the external cores become a continuous ring.
		Our results might help improve predictions for crosstalk in telecommunications or precision in sensing applications. 
	\end{abstract}
	
	\maketitle
%--------------------------------------------------------------------------------
	
%%%%% -*- -*- -*- -*- -*- -*- -*- -*- -*- -*- -*- -*- -*- -*- -*- -*- -*-
\section{Introduction}
		Multicore optical fibers are devices of great importance.
	In telecommunications, they increase transmission bandwidth via space-division multiplexing \cite{Richardson2013} by parallel transmission of, either, signals though different uncoupled cores \cite{Zhu2010,Takara2012,Hu2018}, or normal modes arising from coupled cores \cite{Xia2011,Arik2013}. 
	In sensing, they provide a powerful platform for detection due to the high sensibility of the inter-core couplings to external conditions of the fiber; for example, sensors for curvature, refractive index and temperature have been experimentally demonstrated \cite{Kim2009,Moore2012,SalcedaDelgado2015,Zhao2016,MayArrioja2017a,MayArrioja2017b}. 
	Both in telecommunication and sensing applications, it is crucial to describe the inter-core cross-talk as accurately as possible. 
	
	In multicore fibers, when cores are not excessively close to each other, different cores are coupled by the evanescent tails of their individual modes. 
	This allows the use of coupled-mode theory to approximate electromagnetic field propagation by a linear combination of the individual core modes \cite{Snyder1972,Snyder1988,Huang1994},
	\begin{equation}
	-i \frac{\mathrm{d}}{\mathrm{d}z} \vec{\mathcal{E}}(z) = M \cdot \vec{\mathcal{E}}(z),
	\end{equation}
	where the vector $\vec{\mathcal{E}}(z)$ collects the amplitudes for each individual core mode and $M$ is the mode-coupling matrix that stores the information about effective propagation constanst for the individual core modes in the diagonal and the effective couplings between these modes in the off-diagonal elements.
	The normal modes of the multicore structure are determined by the eigenvectors of the mode-coupling matrix
	and their propagation constants are the corresponding eigenvalues. 
	
	\begin{figure}[htbp]
		\centering
		\fbox{\includegraphics{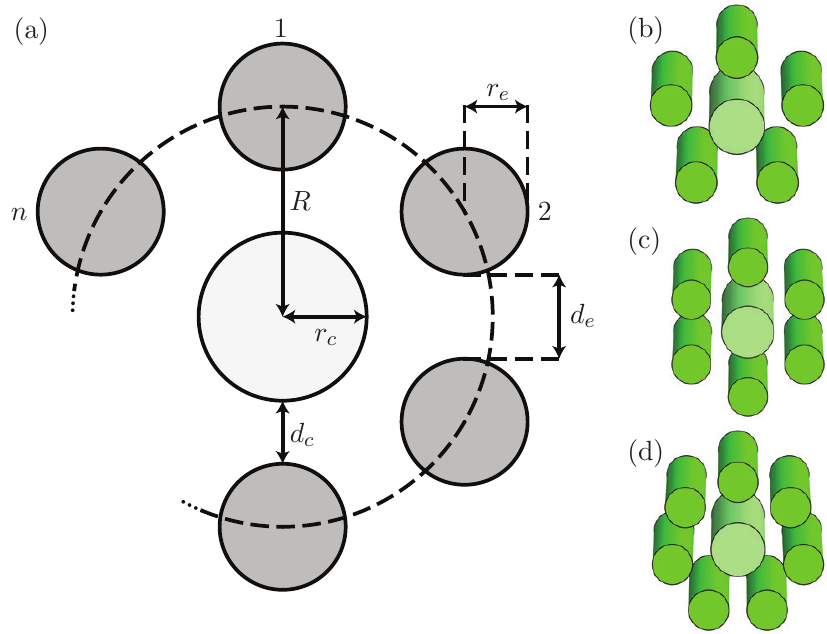}}
		\caption{
			(a) Array of $n$ identical cores distributed in a circle of radius $R$ around a, possibly-different, central core. 
			The cores in the circle have radius $r_e$ and the central core has radius $r_c$. 
			(b)--(d) show the array when the circle has five, six, and seven cores respectively.
		}
		\label{Fig:1:Structure}
	\end{figure}
	
	We focus on a multicore optical fiber with an underlying cyclic symmetry. 
	It is composed by $n$ identical cores evenly distributed on a ring around a possibly-different central core, Fig.~\ref{Fig:1:Structure}. 
	This structure, both with and without the central core, has been studied before using nearest-neighbor coupled-mode theory. 
	In a seminal paper \cite{Snyder1972}, Snyder reduced the problem to that of two coupled effective modes, one with nonzero amplitudes in each of the outer cores and another with nonzero amplitude in just the central core. 
	Yamashita \textit{et al.} used symmetry considerations and Maxwell's equations to study the circular array without a central core \cite{Yamashita1985}.
	Kishi \textit{et al.} followed this approach to study particular cases of the circular array with a central core \cite{Kishi1986,Kishi1988}. 
	The nearest-neighbor coupled-mode theory solution for the circular array is well-known \cite{Schmidt1991,Hudgings2000}. 
	It has been used to study the commensurability of propagation constants \cite{Rubenchik2013} and the stability of nonlinear generalizations \cite{Hizanidis2004,Hizanidis2006,Radosavljevi2015}.
	In particular, this approximation shows good agreement between theory and experiment for the seven-core array \cite{Mortimore1991,Takenaga2010,Koshiba2011,Chan2012,Chekhovskoy2017,Hossain2017}. 
	However, longer propagation distances and higher sensing precision requires going beyond nearest-neighbor interactions \cite{Kishi1988}.
	
	Group-theoretical techniques are a powerful tool to describe electromagnetic field mode amplitude propagation in waveguide arrays with underlying symmetries \cite{Vance1996,RodriguezLara2015,Vergara2015,Huerta2016,NodalStevens2018,RodriguezLara2018}.
	Here, we use the cyclic symmetry of this annular multicore fiber to account for the effective coupling between any pair of cores.
	We focus on the normal modes with vertical polarization but the results can be extended for other polarization modes. 
	In the following, we will assume that each core is a single-mode optical fiber supporting an $LP_{01}$ mode.
	We first find the normal modes for the outer cores without the central core using the discrete Fourier matrix. 
	Then, we include the central core and show that only one of the discrete Fourier modes couples to it.
	Finally, we find the normal modes of the whole fiber and provide their effective propagation constants. 
	For the sake of comparison, we use finite element simulations to confirm our higher-order neighbor results.	
	
	%%%%% -*- -*- -*- -*- -*- -*- -*- -*- -*- -*- -*- -*- -*- -*- -*- -*- -*-
   \section{Outer ring cores}
	First, we focus on the outer ring composed by $n$ identical single-mode fiber cores evenly placed in a circular array.
	We label the cores from $1$ to $n$ in clockwise order, Fig. \ref{Fig:1:Structure}. 
	Since all the cores are identical, we label the effective propagation constant for every single core mode as $\beta$.
	We call the effective $k$-th neigbor coupling $g_{k}$ with $k=1,2,\ldots, \lfloor n/2 \rfloor$; the function $\lfloor x \rfloor$ yields the greatest integer less than or equal to the argument $x$. 
	The coupling between a pair of cores decreases with the distance, $g_1 > g_2 > \ldots > g_{\lfloor n/2 \rfloor}$. 
	We can write the elements of the mode-coupling matrix in the following manner, 
	\begin{eqnarray}
	\left[ M \right]_{p,q} = \left\{ \begin{array}{lll} 
	g_{k} 	& q = p + n-k, 	& k =1,2, \ldots, \lfloor \frac{n-1}{2} \rfloor, 	\\ 
	g_{k} 	& q = p + k, 	& k =1,2, \ldots, \lfloor \frac{n}{2} \rfloor, 		\\ 
	\beta 	& q = p, 		& 									\\ 
	g_{k} 	& p = q + k, 	& k =1,2, \ldots, \lfloor \frac{n}{2} \rfloor, 		\\ 
	g_{k} 	& p = q +n-k, 	& k =1,2, \ldots, \lfloor \frac{n-1}{2} \rfloor.
	\end{array} \right. 
	\end{eqnarray}
	The first and last upper and lower diagonals in the mode-coupling matrix contain all the first neighbor couplings, $g_{1}$, and so on.
	The cyclic symmetry of the system produces shifts in the columns of the coupled-mode matrix. 
	This symmetry is embodied by the discrete cyclic group with $n$ elements, $\mathbb{Z}_{n}$. 
	We have used in the past the fact that both the representation of the generator of $\mathbb{Z}_{n}$ and the coupled-mode matrix are diagonalized by the Fourier matrix \cite{NodalStevens2018},
	\begin{eqnarray}\label{eq:FourierMatrix}
	\left[ F \right]_{p,q} = \frac{1}{\sqrt{n}} e^{i \frac{2\pi}{n} (p-1) (q-1) }.
	\end{eqnarray} 
	This approach yields the propagation constants for the normal modes of the circular array,
	\begin{eqnarray}\label{Eq:CircularArray:PropagationConstants}
	\lambda_j
	= 	\beta + 
	\left\{ \begin{array}{ll}
	2 \sum\limits_{k=1}^{m-1} \left\{ g_{k} \, \cos\left[ \frac{\pi}{m} (j-1) k \right] \right\} + g_{m} (-1)^{j-1}, & n= 2m, \\
	2 \sum\limits_{k=1}^{m}	\left\{ g_{k} \, \cos\left[ \frac{2\pi}{2m+1} (j-1) k \right] \right\}, & n=2m+1. \\
	\end{array} \right.
	\end{eqnarray}
	In the first-neighbors approximation, where $g_k \to 0$ for $k \geq 2$, these values match previous results \cite{Schmidt1991,Rubenchik2013}.
	The first mode propagation constant, $\lambda_1$, is always the largest value because all the arguments in the cosines are zero. 
	Its corresponding supermode has equal field amplitudes in all the cores. 
	Additionally, there is multiplicity in the propagation constants. 
	The second propagation constant is equal to the $n$-th one, $\lambda_2=\lambda_n$, the third is equal to the $(n-1)$-th one, $\lambda_3=\lambda_{n-1}$, and so on. 
	When $n$ is even, there are $n/2 - 1$ duplicated propagation constants and two non-duplicated ones, $\lambda_1$ and $\lambda_{n/2 + 1}$. 
	When $n$ is odd, there are $(n-1)/2$ duplicated and a single non-duplicated, $\lambda_1$.
	
	The normal modes, or supermodes, are given by the action of the discrete Fourier matrix on the standard orthonormal basis \cite{NodalStevens2018}.
	The mode corresponding to the $j$-th effective propagation constant is the $j$-th column of the conjugate transpose of the discrete Fourier matrix, 
	\begin{equation}
	\vec{u}_j = F^{\dagger} \cdot \hat{e}_j,
	\end{equation}
	where $\hat{e}_j$ is $j$-th vector of the standard orthonormal basis. 
	This allows us to realize that the normal modes are independent of the coupling and propagation constants, Eq.~(\ref{eq:FourierMatrix}).
	They form a complete orthonormal basis, thus the duplicity in propagation constants is not a degeneracy.

	\section{Full multicore fiber}
	Now, we add the central core by assigning an effective propagation constant $\beta_c$ to it and an effective coupling $g_c$ between it and each other core in the outer ring. 
	The mode-coupling matrix for the whole system can be written as a block matrix,
	\begin{equation}
	M_c 
	= 
	\left(
	\begin{array}{c|c}
	M
	& \begin{array}{c} g_c \\ \vdots \\ g_c \end{array}
	\\
	\hline
	\begin{array}{ccc} g_c & \cdots & g_c \end{array}
	& \beta_c
	\end{array}
	\right).
	\end{equation}
	This matrix can be diagonalized by extending our procedure above. 
	This leads to an effective coupling, with strength $\sqrt{n} g_c$, between the central single core mode
	and the first supermode of the external cores, 
	\begin{equation}
	D_{c}
	=
	\left(
	\begin{array}{cccc|c}
	\lambda_1		& 0				& \ldots		& 0				& g_c \sqrt{n}		\\
	0				& \lambda_2		& \ddots		& \vdots		& 0					\\
	\vdots			& \ddots		& \ddots		& 0				& \vdots			\\
	0				& \ldots		& 0				& \lambda_n		& 0					\\
	\hline
	g_c \sqrt{n}	& 0				& \cdots		& 0				& \beta_c
	\end{array}
	\right).
	\end{equation}
	Snyder discussed the coupling between these two modes \cite{Snyder1972}. 
	The rest of the outer ring supermodes do not couple to the central core and, therefore, remain normal modes when the central core is added. 
	It is straightforward to diagonalize the effective two-by-two real-symmetric mode-coupling matrix $D_{c}$ and obtain the propagation constants,
	\begin{eqnarray}\label{Eq:PropagationConstantsWithCentral}
	&&\lambda_{\pm} = \left( \lambda_1 + \beta_c \pm \sqrt{  \left( \lambda_1 - \beta_c \right)^2  +  4 \, g_c^2 \, n  }, \right)/2,	\nonumber \\
	&&\lambda_2, \ldots, \lambda_n,
	\end{eqnarray}
	for the supermodes of the whole multicore fiber. 
	The propagation constant $\lambda_{+}$ is the largest, but $\lambda_{-}$ is not necessarily the smallest. 
	The supermodes with propagation constants $\lambda_{+}$ and $\lambda_{-}$ have nonzero complex field amplitudes in the central core.
	The mode associated with $\lambda_{+}$ has field amplitudes with identical phases in all cores. 
	That associated with $\lambda_{-}$ has a $\pi$ phase difference between the central core and the outer field amplitudes. 
	We will call these symmetric and anti-symmetric phase modes, 
	\begin{eqnarray}
	\vec{v}_{+} &=&
	\left(\begin{array}{c} \vec{u}_{1} \\ 0 \end{array}\right) \cos \theta 
	+\left(\begin{array}{c} \vec{0}_{n} \\ 1 \end{array}\right) \sin \theta, \nonumber \\
	\vec{v}_{-} &=&
	- \left(\begin{array}{c} \vec{u}_{1} \\ 0 \end{array}\right) \sin \theta 
	+\left(\begin{array}{c} \vec{0}_{n} \\ 1 \end{array}\right) \cos \theta,
	\end{eqnarray}
	where the $n$-dimensional vectors $\vec{u}_{1}$ and $\vec{0}_{n}$ are the first supermode of the external ring and the zero vector, in that order. 
	The mixing angle,
	\begin{equation}
	\tan \theta = \frac{2\sqrt{n}g_{c}}{\lambda_{1}-\beta_{c}+\sqrt{(\lambda_{1}-\beta_{c})^2+4n g_{c}^2}},
	\end{equation}
	allows us to control the total field intensity distribution in the outer ring and the central core. 
	For example, if the system is designed to fulfill $\lambda_{1} = \beta_{c}$, the total power in the outer cores will be identical to the power in the central core. 
	
	\begin{figure}[htbp]
		\centering
		\fbox{\includegraphics{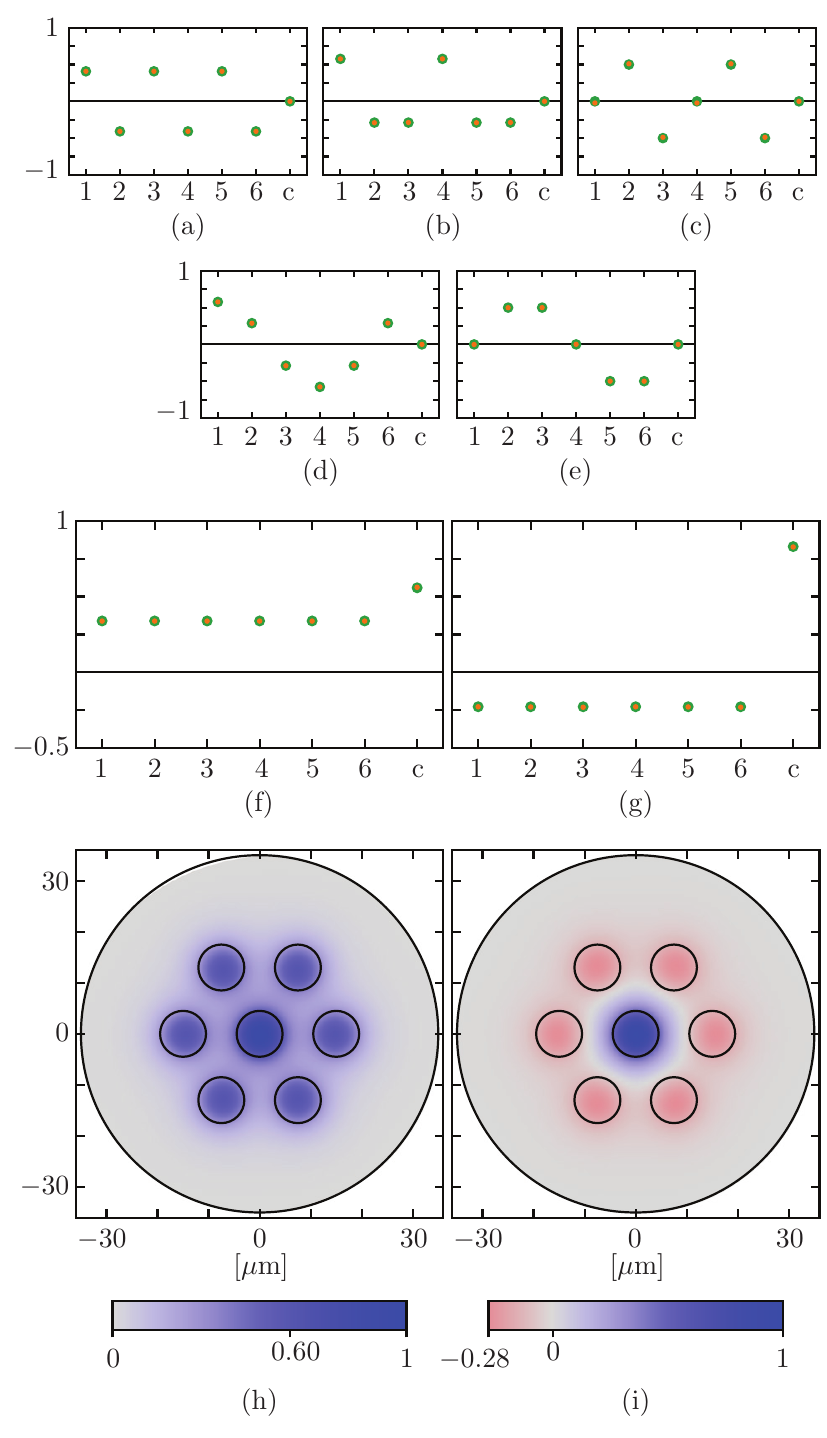}}
		\caption{
			(a)--(g) Comparison of analytic coupled-mode theory (small light orange dots) versus numerical finite element (large dark green dots) normal modes for a multicore fiber composed by seven identical cores; fiber parameters can be found in the text. 
			(a)--(e) Normal modes that do not mix with the central core and (f)--(g) modes that mix with the central core. 
			The propagation constants associated to the normal modes shown in (f) and (g) are $\lambda_{+}$ and $\lambda_{-}$, respectively.
			In these figures, the horizontal axis is the core number, where $c$ labels the central core and the vertical axis is the real part of the average electric field in each core. 
			Figures (h) and (i) display the real part of the electric field in a cross-section of the waveguide, where the electric field is normalized to its maximum value in the array, for normal modes associated to $\lambda_{+}$ and $\lambda_{-}$, in that order.
		}
		\label{Fig:2:Modes}
	\end{figure}
	
	We use numerical finite element simulation to validate our coupled-mode theory analysis. 
	Figure \ref{Fig:2:Modes} shows the results for a multicore fiber composed by identical cores and cladding with refractive indices $n_{e} = n_{c} = 1.4479$ and $n_{cl} = 1.4440$, in that order. 
	The radii of all cores are identical too, $r_{e} = r_{c} = 4.5~\mu\mathrm{m}$, as are the distances between each external and central core, $R=15~\mu\mathrm{m}$.
	These parameters are typical in Silica multicore fibers at telecomm wavelength, $\lambda = 1550~\mathrm{nm}$ \cite{MayArrioja2017a}.
	We assume a cladding radius of $35~\mu\mathrm{m}$. 
	The cores support a $\mathrm{LP}_{01}$ mode that can be approximated to a Gaussian with waist $\omega_{0} = 5.5564~\mu\mathrm{m}$.
	Coupled-mode theory translates these parameters into the following effective propagation constants and couplings, $\beta = \beta_{c} \approx 5.8606 \times 10^{6} ~\mathrm{rad}/\mathrm{m}$ and $g_{1} \approx  g_{c} = 310.00 ~\mathrm{rad}/\mathrm{m}$, respectively.
	We use our analytic result and the finite element simulation to fit and recover numerical values of $\beta = 5.8598\times 10^{6}~ \mathrm{rad}/\mathrm{m}$ and $g_{1} = g_{c} = 312.44~ \textrm{rad}/ \mathrm{m}$. 
	That is a $0.0137 \%$ and $0.78 \%$ relative difference with respect to the numerically fitted values, in that order.
	
\begin{figure}[htbp]
		\centering
		\fbox{\includegraphics{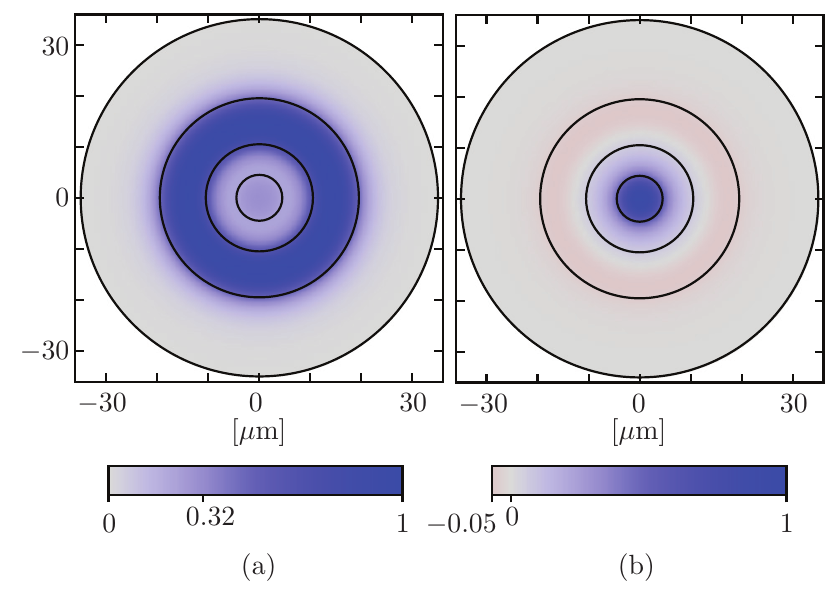}}
		\caption{
			Mode profiles for the multicore structure with a continuous ring around a central core with parameters identical to those in Fig. \ref{Fig:2:Modes}.
		}
		\label{Fig:3:RingModes}
	\end{figure}

	For the sake of curiousity, we study a continuous case where a ring core surrounds the central core. 
	A type of this concentric ring fiber has been used to experimentaly measure the Abraham force of light over a liquid contained in a central hollow core \cite{Choi2017}. 
	We expected the symmetric and anti-symmetric phase modes to survive with effective propagation constants in the outer ring and central core given by $\beta_{R}$ and $\beta_{c}$, in that order, and a finite coupling where $g_{c} \sqrt{n} \rightarrow g_{R}$. 
	This yields an effective two-mode coupled array with propagation constants $\lambda_{(R,\pm)} = \left( \beta_{R} + \beta_{c} \pm \sqrt{ \left( \beta_{R} - \beta_{c} \right)^{2} + 4 g_{R}^2} \right)/2$ to good agreement with the finite element simulation providing $\lambda_{(R,+)} = 5.8651\times 10^{6}~ \mathrm{rad}/\mathrm{m}$ and $\lambda_{(R,-)}=5.8598\times 10^{6}~ \mathrm{rad}/\mathrm{m}$, Fig.~\ref{Fig:3:RingModes}.
	We fit these numerical results to the analytic approximation to obtain an effective coupling between outer ring and central core, $g_{R}=1401~\mathrm{rad}/\mathrm{m}$. 
	This fit uses an effective propagation constant for the outer ring mode that comes from a finite element simulation of the ring alone, 
	$\beta_{R} = 5.8651 \times 10^{6} ~\mathrm{rad}/\mathrm{m}$, 
	and the propagation constant for the central core that is identical with the discrete case, $\beta_{c} = 5.8606 \times 10^{6} ~\mathrm{rad}/\mathrm{m}$.
	
	%%%%% -*- -*- -*- -*- -*- -*- -*- -*- -*- -*- -*- -*- -*- -*- -*- -*- -*-
	\section{Conclusions}
	In summary, we provided a symmetry-based analysis for a multicore fiber composed by $n$ evenly spaced cores around a central core. 
	The cyclic group $\mathbb{Z}_{n}$ allowed us to find the normal modes and their propagation constants considering all inter-core couplings. 
	First, we used the discrete Fourier matrix to resolve the normal modes of the outer cores and their propagation constants; none of these supermodes depend on the parameters of the system.
	Then, we studied how these outer core modes couple to the central core and confirmed Snyder's seminal result \cite{Snyder1972}: 
	only the supermode composed by equal complex field amplitudes in all the outer cores couples to the central core.
	The rest $n-1$ outer core supermodes, whose field distribution does not depend on the parameters of the system, remain uncoupled to the single central core mode. 
	
	We used finite element analysis to verify our theoretical results for a fiber where all cores are identical to a good agreement; the relative differences between analytic and numerical results were of the order of $0.01 \%$ and $0.78 \%$ of the numerical value for the effective propagation constant of each core and the coupling between cores, in that order. 
	Our symmetry-based analysis provides a description that might help improve the prediction of crosstalk through propagation in multicore fibers for telecomm and sensing applications.

	% \section{Funding}
	\textbf{Funding.} 
	CONACYT, C\'{a}tedra Grupal \#551; 
	CONACYT, Consorcio en \'Optica Aplicada, FORDECYT \#296355. 
	Academia Mexicana de Ciencias through Verano de la Investigac\'on Cient\'ifica.
	
	%\textbf{Acknowledgment.}
	%S. M. T. R. acknowleges financial support from the Academia Mexicana de Ciencias through the program Verano de la Investigac\'on Cient\'ifica.
	%B.M.R.-L. thanks the Photonics and Mathematical Optics Group at Tecnologico de Monterrey.

% \section{References}
% Bibliography
%-%-%-%-%-%-%-%-%-%-%-%-%-%-%-%-%-%-%-%-%-%-%-%-\bibliography{ref-NecklaceWCentral}

\begin{thebibliography}{37}%
	\makeatletter
	\providecommand \@ifxundefined [1]{%
		\@ifx{#1\undefined}
	}%
	\providecommand \@ifnum [1]{%
		\ifnum #1\expandafter \@firstoftwo
		\else \expandafter \@secondoftwo
		\fi
	}%
	\providecommand \@ifx [1]{%
		\ifx #1\expandafter \@firstoftwo
		\else \expandafter \@secondoftwo
		\fi
	}%
	\providecommand \natexlab [1]{#1}%
	\providecommand \enquote  [1]{``#1''}%
	\providecommand \bibnamefont  [1]{#1}%
	\providecommand \bibfnamefont [1]{#1}%
	\providecommand \citenamefont [1]{#1}%
	\providecommand \href@noop [0]{\@secondoftwo}%
	\providecommand \href [0]{\begingroup \@sanitize@url \@href}%
	\providecommand \@href[1]{\@@startlink{#1}\@@href}%
	\providecommand \@@href[1]{\endgroup#1\@@endlink}%
	\providecommand \@sanitize@url [0]{\catcode `\\12\catcode `\$12\catcode
		`\&12\catcode `\#12\catcode `\^12\catcode `\_12\catcode `\%12\relax}%
	\providecommand \@@startlink[1]{}%
	\providecommand \@@endlink[0]{}%
	\providecommand \url  [0]{\begingroup\@sanitize@url \@url }%
	\providecommand \@url [1]{\endgroup\@href {#1}{\urlprefix }}%
	\providecommand \urlprefix  [0]{URL }%
	\providecommand \Eprint [0]{\href }%
	\providecommand \doibase [0]{http://dx.doi.org/}%
	\providecommand \selectlanguage [0]{\@gobble}%
	\providecommand \bibinfo  [0]{\@secondoftwo}%
	\providecommand \bibfield  [0]{\@secondoftwo}%
	\providecommand \translation [1]{[#1]}%
	\providecommand \BibitemOpen [0]{}%
	\providecommand \bibitemStop [0]{}%
	\providecommand \bibitemNoStop [0]{.\EOS\space}%
	\providecommand \EOS [0]{\spacefactor3000\relax}%
	\providecommand \BibitemShut  [1]{\csname bibitem#1\endcsname}%
	\let\auto@bib@innerbib\@empty
	%</preamble>
	\bibitem [{\citenamefont {Richardson}\ \emph {et~al.}(2013)\citenamefont
		{Richardson}, \citenamefont {Fini},\ and\ \citenamefont
		{Nelson}}]{Richardson2013}%
	\BibitemOpen
	\bibfield  {author} {\bibinfo {author} {\bibfnamefont {D.~J.}\ \bibnamefont
			{Richardson}}, \bibinfo {author} {\bibfnamefont {J.~M.}\ \bibnamefont
			{Fini}}, \ and\ \bibinfo {author} {\bibfnamefont {L.~E.}\ \bibnamefont
			{Nelson}},\ }\bibfield  {title} {\enquote {\bibinfo {title} {Space-division
				multiplexing in optical fibres},}\ }\href {\doibase 10.1038/nphoton.2013.94}
	{\bibfield  {journal} {\bibinfo  {journal} {Nature Photonics}\ }\textbf
		{\bibinfo {volume} {7}},\ \bibinfo {pages} {354} (\bibinfo {year}
		{2013})}\BibitemShut {NoStop}%
	\bibitem [{\citenamefont {Zhu}\ \emph {et~al.}(2010)\citenamefont {Zhu},
		\citenamefont {Taunay}, \citenamefont {Yan}, \citenamefont {Fini},
		\citenamefont {Fishteyn}, \citenamefont {Monberg},\ and\ \citenamefont
		{Dimarcello}}]{Zhu2010}%
	\BibitemOpen
	\bibfield  {author} {\bibinfo {author} {\bibfnamefont {B.}~\bibnamefont
			{Zhu}}, \bibinfo {author} {\bibfnamefont {T.~F.}\ \bibnamefont {Taunay}},
		\bibinfo {author} {\bibfnamefont {M.~F.}\ \bibnamefont {Yan}}, \bibinfo
		{author} {\bibfnamefont {J.~M.}\ \bibnamefont {Fini}}, \bibinfo {author}
		{\bibfnamefont {M.}~\bibnamefont {Fishteyn}}, \bibinfo {author}
		{\bibfnamefont {E.~M.}\ \bibnamefont {Monberg}}, \ and\ \bibinfo {author}
		{\bibfnamefont {F.~V.}\ \bibnamefont {Dimarcello}},\ }\bibfield  {title}
	{\enquote {\bibinfo {title} {Seven-core multicore fiber transmissions for
				passive optical network},}\ }\href {\doibase 10.1364/OE.18.011117} {\bibfield
		{journal} {\bibinfo  {journal} {Opt. Express}\ }\textbf {\bibinfo {volume}
			{18}},\ \bibinfo {pages} {11117--11122} (\bibinfo {year} {2010})}\BibitemShut
	{NoStop}%
	\bibitem [{\citenamefont {Takara}\ \emph {et~al.}(2012)\citenamefont {Takara},
		\citenamefont {Sano}, \citenamefont {Kobayashi}, \citenamefont {Kubota},
		\citenamefont {Kawakami}, \citenamefont {Matsuura}, \citenamefont {Miyamoto},
		\citenamefont {Abe}, \citenamefont {Ono}, \citenamefont {Shikama},
		\citenamefont {Goto}, \citenamefont {Tsujikawa}, \citenamefont {Sasaki},
		\citenamefont {Ishida}, \citenamefont {Takenaga}, \citenamefont {Matsuo},
		\citenamefont {Saitoh}, \citenamefont {Koshiba},\ and\ \citenamefont
		{Morioka}}]{Takara2012}%
	\BibitemOpen
	\bibfield  {author} {\bibinfo {author} {\bibfnamefont {H.}~\bibnamefont
			{Takara}}, \bibinfo {author} {\bibfnamefont {A.}~\bibnamefont {Sano}},
		\bibinfo {author} {\bibfnamefont {T.}~\bibnamefont {Kobayashi}}, \bibinfo
		{author} {\bibfnamefont {H.}~\bibnamefont {Kubota}}, \bibinfo {author}
		{\bibfnamefont {H.}~\bibnamefont {Kawakami}}, \bibinfo {author}
		{\bibfnamefont {A.}~\bibnamefont {Matsuura}}, \bibinfo {author}
		{\bibfnamefont {Y.}~\bibnamefont {Miyamoto}}, \bibinfo {author}
		{\bibfnamefont {Y.}~\bibnamefont {Abe}}, \bibinfo {author} {\bibfnamefont
			{H.}~\bibnamefont {Ono}}, \bibinfo {author} {\bibfnamefont {K.}~\bibnamefont
			{Shikama}}, \bibinfo {author} {\bibfnamefont {Y.}~\bibnamefont {Goto}},
		\bibinfo {author} {\bibfnamefont {K.}~\bibnamefont {Tsujikawa}}, \bibinfo
		{author} {\bibfnamefont {Y.}~\bibnamefont {Sasaki}}, \bibinfo {author}
		{\bibfnamefont {I.}~\bibnamefont {Ishida}}, \bibinfo {author} {\bibfnamefont
			{K.}~\bibnamefont {Takenaga}}, \bibinfo {author} {\bibfnamefont
			{S.}~\bibnamefont {Matsuo}}, \bibinfo {author} {\bibfnamefont
			{K.}~\bibnamefont {Saitoh}}, \bibinfo {author} {\bibfnamefont
			{M.}~\bibnamefont {Koshiba}}, \ and\ \bibinfo {author} {\bibfnamefont
			{T.}~\bibnamefont {Morioka}},\ }\bibfield  {title} {\enquote {\bibinfo
			{title} {1.01-pb/s (12 sdm/222 wdm/456 gb/s) crosstalk-managed transmission
				with 91.4-b/s/hz aggregate spectral efficiency},}\ }in\ \href {\doibase
		10.1364/ECEOC.2012.Th.3.C.1} {\emph {\bibinfo {booktitle} {European
				Conference and Exhibition on Optical Communication}}}\ (\bibinfo {year}
	{2012})\ p.\ \bibinfo {pages} {Th.3.C.1}\BibitemShut {NoStop}%
	\bibitem [{\citenamefont {Hu}\ \emph {et~al.}(2018)\citenamefont {Hu},
		\citenamefont {Da~Ros}, \citenamefont {Pu}, \citenamefont {Ye}, \citenamefont
		{Ingerslev}, \citenamefont {Porto~da Silva}, \citenamefont {Nooruzzaman},
		\citenamefont {Amma}, \citenamefont {Sasaki}, \citenamefont {Mizuno},
		\citenamefont {Miyamoto}, \citenamefont {Ottaviano}, \citenamefont
		{Semenova}, \citenamefont {Guan}, \citenamefont {Zibar}, \citenamefont
		{Galili}, \citenamefont {Yvind}, \citenamefont {Morioka},\ and\ \citenamefont
		{Oxenl{\o}we}}]{Hu2018}%
	\BibitemOpen
	\bibfield  {author} {\bibinfo {author} {\bibfnamefont {Hao}\ \bibnamefont
			{Hu}}, \bibinfo {author} {\bibfnamefont {Francesco}\ \bibnamefont {Da~Ros}},
		\bibinfo {author} {\bibfnamefont {Minhao}\ \bibnamefont {Pu}}, \bibinfo
		{author} {\bibfnamefont {Feihong}\ \bibnamefont {Ye}}, \bibinfo {author}
		{\bibfnamefont {Kasper}\ \bibnamefont {Ingerslev}}, \bibinfo {author}
		{\bibfnamefont {Edson}\ \bibnamefont {Porto~da Silva}}, \bibinfo {author}
		{\bibfnamefont {Md}~\bibnamefont {Nooruzzaman}}, \bibinfo {author}
		{\bibfnamefont {Yoshimichi}\ \bibnamefont {Amma}}, \bibinfo {author}
		{\bibfnamefont {Yusuke}\ \bibnamefont {Sasaki}}, \bibinfo {author}
		{\bibfnamefont {Takayuki}\ \bibnamefont {Mizuno}}, \bibinfo {author}
		{\bibfnamefont {Yutaka}\ \bibnamefont {Miyamoto}}, \bibinfo {author}
		{\bibfnamefont {Luisa}\ \bibnamefont {Ottaviano}}, \bibinfo {author}
		{\bibfnamefont {Elizaveta}\ \bibnamefont {Semenova}}, \bibinfo {author}
		{\bibfnamefont {Pengyu}\ \bibnamefont {Guan}}, \bibinfo {author}
		{\bibfnamefont {Darko}\ \bibnamefont {Zibar}}, \bibinfo {author}
		{\bibfnamefont {Michael}\ \bibnamefont {Galili}}, \bibinfo {author}
		{\bibfnamefont {Kresten}\ \bibnamefont {Yvind}}, \bibinfo {author}
		{\bibfnamefont {Toshio}\ \bibnamefont {Morioka}}, \ and\ \bibinfo {author}
		{\bibfnamefont {Leif~K.}\ \bibnamefont {Oxenl{\o}we}},\ }\bibfield  {title}
	{\enquote {\bibinfo {title} {Single-source chip-based frequency comb enabling
				extreme parallel data transmission},}\ }\href {\doibase
		10.1038/s41566-018-0205-5} {\bibfield  {journal} {\bibinfo  {journal} {Nature
				Photonics}\ }\textbf {\bibinfo {volume} {12}},\ \bibinfo {pages} {469}
		(\bibinfo {year} {2018})}\BibitemShut {NoStop}%
	\bibitem [{\citenamefont {Xia}\ \emph {et~al.}(2011)\citenamefont {Xia},
		\citenamefont {Bai}, \citenamefont {Ozdur}, \citenamefont {Zhou},\ and\
		\citenamefont {Li}}]{Xia2011}%
	\BibitemOpen
	\bibfield  {author} {\bibinfo {author} {\bibfnamefont {Cen}\ \bibnamefont
			{Xia}}, \bibinfo {author} {\bibfnamefont {Neng}\ \bibnamefont {Bai}},
		\bibinfo {author} {\bibfnamefont {Ibrahim}\ \bibnamefont {Ozdur}}, \bibinfo
		{author} {\bibfnamefont {Xiang}\ \bibnamefont {Zhou}}, \ and\ \bibinfo
		{author} {\bibfnamefont {Guifang}\ \bibnamefont {Li}},\ }\bibfield  {title}
	{\enquote {\bibinfo {title} {Supermodes for optical transmission},}\ }\href
	{\doibase 10.1364/OE.19.016653} {\bibfield  {journal} {\bibinfo  {journal}
			{Optics Express}\ }\textbf {\bibinfo {volume} {19}},\ \bibinfo {pages}
		{16653} (\bibinfo {year} {2011})}\BibitemShut {NoStop}%
	\bibitem [{\citenamefont {Ar{\i}k}\ and\ \citenamefont
		{Kahn}(2013)}]{Arik2013}%
	\BibitemOpen
	\bibfield  {author} {\bibinfo {author} {\bibfnamefont {Sercan~\"O.}\
			\bibnamefont {Ar{\i}k}}\ and\ \bibinfo {author} {\bibfnamefont {Joseph~M.}\
			\bibnamefont {Kahn}},\ }\bibfield  {title} {\enquote {\bibinfo {title}
			{Coupled-core multi-core fibers for spatial multiplexing},}\ }\href {\doibase
		10.1109/LPT.2013.2280897} {\bibfield  {journal} {\bibinfo  {journal} {IEEE
				Photonics Technology Letters}\ }\textbf {\bibinfo {volume} {25}},\ \bibinfo
		{pages} {2054} (\bibinfo {year} {2013})}\BibitemShut {NoStop}%
	\bibitem [{\citenamefont {Kim}\ \emph {et~al.}(2009)\citenamefont {Kim},
		\citenamefont {Kim}, \citenamefont {Cui},\ and\ \citenamefont
		{Chung}}]{Kim2009}%
	\BibitemOpen
	\bibfield  {author} {\bibinfo {author} {\bibfnamefont {Bongkyun}\
			\bibnamefont {Kim}}, \bibinfo {author} {\bibfnamefont {Tae-Hoon}\
			\bibnamefont {Kim}}, \bibinfo {author} {\bibfnamefont {Long}\ \bibnamefont
			{Cui}}, \ and\ \bibinfo {author} {\bibfnamefont {Youngjoo}\ \bibnamefont
			{Chung}},\ }\bibfield  {title} {\enquote {\bibinfo {title} {Twin core
				photonic crystal fiber for in-line mach-zehnder interferometric sensing
				applications},}\ }\href {\doibase 10.1364/OE.17.015502} {\bibfield  {journal}
		{\bibinfo  {journal} {Opt. Express}\ }\textbf {\bibinfo {volume} {17}},\
		\bibinfo {pages} {15502} (\bibinfo {year} {2009})}\BibitemShut {NoStop}%
	\bibitem [{\citenamefont {Moore}\ and\ \citenamefont
		{Rogge}(2012)}]{Moore2012}%
	\BibitemOpen
	\bibfield  {author} {\bibinfo {author} {\bibfnamefont {Jason~P.}\
			\bibnamefont {Moore}}\ and\ \bibinfo {author} {\bibfnamefont {Matthew~D.}\
			\bibnamefont {Rogge}},\ }\bibfield  {title} {\enquote {\bibinfo {title}
			{Shape sensing using multi-core fiber optic cable and parametric curve
				solutions},}\ }\href {\doibase 10.1364/OE.20.002967} {\bibfield  {journal}
		{\bibinfo  {journal} {Opt. Express}\ }\textbf {\bibinfo {volume} {20}},\
		\bibinfo {pages} {2967} (\bibinfo {year} {2012})}\BibitemShut {NoStop}%
	\bibitem [{\citenamefont {Salceda-Delgado}\ \emph {et~al.}(2015)\citenamefont
		{Salceda-Delgado}, \citenamefont {Newkirk}, \citenamefont {Antonio-Lopez},
		\citenamefont {Martinez-Rios}, \citenamefont {Sch\"{u}lzgen},\ and\
		\citenamefont {Correa}}]{SalcedaDelgado2015}%
	\BibitemOpen
	\bibfield  {author} {\bibinfo {author} {\bibfnamefont {G.}~\bibnamefont
			{Salceda-Delgado}}, \bibinfo {author} {\bibfnamefont {A.~Van}\ \bibnamefont
			{Newkirk}}, \bibinfo {author} {\bibfnamefont {J.~E.}\ \bibnamefont
			{Antonio-Lopez}}, \bibinfo {author} {\bibfnamefont {A.}~\bibnamefont
			{Martinez-Rios}}, \bibinfo {author} {\bibfnamefont {A.}~\bibnamefont
			{Sch\"{u}lzgen}}, \ and\ \bibinfo {author} {\bibfnamefont {R.~Amezcua}\
			\bibnamefont {Correa}},\ }\bibfield  {title} {\enquote {\bibinfo {title}
			{Compact fiber-optic curvature sensor based on super-mode interference in a
				seven-core fiber},}\ }\href {\doibase 10.1364/OL.40.001468} {\bibfield
		{journal} {\bibinfo  {journal} {Opt. Lett.}\ }\textbf {\bibinfo {volume}
			{40}},\ \bibinfo {pages} {1468--1471} (\bibinfo {year} {2015})}\BibitemShut
	{NoStop}%
	\bibitem [{\citenamefont {Zhao}\ \emph {et~al.}(2016)\citenamefont {Zhao},
		\citenamefont {Soto}, \citenamefont {Tang},\ and\ \citenamefont
		{Th\'{e}venaz}}]{Zhao2016}%
	\BibitemOpen
	\bibfield  {author} {\bibinfo {author} {\bibfnamefont {Zhiyong}\ \bibnamefont
			{Zhao}}, \bibinfo {author} {\bibfnamefont {Marcelo~A.}\ \bibnamefont {Soto}},
		\bibinfo {author} {\bibfnamefont {Ming}\ \bibnamefont {Tang}}, \ and\
		\bibinfo {author} {\bibfnamefont {Luc}\ \bibnamefont {Th\'{e}venaz}},\
	}\bibfield  {title} {\enquote {\bibinfo {title} {Distributed shape sensing
				using brillouin scattering in multi-core fibers},}\ }\href {\doibase
		10.1364/OE.24.025211} {\bibfield  {journal} {\bibinfo  {journal} {Opt.
				Express}\ }\textbf {\bibinfo {volume} {24}},\ \bibinfo {pages} {25211}
		(\bibinfo {year} {2016})}\BibitemShut {NoStop}%
	\bibitem [{\citenamefont {May-Arrioja}\ and\ \citenamefont
		{Guzman-Sepulveda}(2017{\natexlab{a}})}]{MayArrioja2017a}%
	\BibitemOpen
	\bibfield  {author} {\bibinfo {author} {\bibfnamefont {Daniel~Alberto}\
			\bibnamefont {May-Arrioja}}\ and\ \bibinfo {author} {\bibfnamefont
			{Jose~Rafael}\ \bibnamefont {Guzman-Sepulveda}},\ }\bibfield  {title}
	{\enquote {\bibinfo {title} {Highly sensitive fiber optic refractive index
				sensor using multicore coupled structures},}\ }\href {\doibase
		10.1109/JLT.2017.2699619} {\bibfield  {journal} {\bibinfo  {journal} {Journal
				of Lightwave Technology}\ }\textbf {\bibinfo {volume} {35}},\ \bibinfo
		{pages} {2695--2701} (\bibinfo {year} {2017}{\natexlab{a}})}\BibitemShut
	{NoStop}%
	\bibitem [{\citenamefont {May-Arrioja}\ and\ \citenamefont
		{Guzman-Sepulveda}(2017{\natexlab{b}})}]{MayArrioja2017b}%
	\BibitemOpen
	\bibfield  {author} {\bibinfo {author} {\bibfnamefont {D.~A.}\ \bibnamefont
			{May-Arrioja}}\ and\ \bibinfo {author} {\bibfnamefont {J.~R.}\ \bibnamefont
			{Guzman-Sepulveda}},\ }\bibfield  {title} {\enquote {\bibinfo {title} {Fiber
				optic sensors based on multicore structures},}\ }in\ \href {\doibase
		10.1007/978-3-319-42625-9_16} {\emph {\bibinfo {booktitle} {Fiber Optic
				Sensors: Current Status and Future Possibilities}}},\ \bibinfo {editor}
	{edited by\ \bibinfo {editor} {\bibfnamefont {Ignacio~R.}\ \bibnamefont
			{Matias}}, \bibinfo {editor} {\bibfnamefont {Satoshi}\ \bibnamefont
			{Ikezawa}}, \ and\ \bibinfo {editor} {\bibfnamefont {Jesus}\ \bibnamefont
			{Corres}}}\ (\bibinfo  {publisher} {Springer International Publishing},\
	\bibinfo {year} {2017})\ pp.\ \bibinfo {pages} {347--371}\BibitemShut
	{NoStop}%
	\bibitem [{\citenamefont {Snyder}(1972)}]{Snyder1972}%
	\BibitemOpen
	\bibfield  {author} {\bibinfo {author} {\bibfnamefont {Allan~W.}\
			\bibnamefont {Snyder}},\ }\bibfield  {title} {\enquote {\bibinfo {title}
			{Coupled-mode theory for optical fibers},}\ }\href {\doibase
		10.1364/JOSA.62.001267} {\bibfield  {journal} {\bibinfo  {journal} {J. Opt.
				Soc. Am.}\ }\textbf {\bibinfo {volume} {62}},\ \bibinfo {pages} {1267--1277}
		(\bibinfo {year} {1972})}\BibitemShut {NoStop}%
	\bibitem [{\citenamefont {Snyder}\ and\ \citenamefont
		{Ankiewicz}(1988)}]{Snyder1988}%
	\BibitemOpen
	\bibfield  {author} {\bibinfo {author} {\bibfnamefont {Allan~W.}\
			\bibnamefont {Snyder}}\ and\ \bibinfo {author} {\bibfnamefont {Adrian}\
			\bibnamefont {Ankiewicz}},\ }\bibfield  {title} {\enquote {\bibinfo {title}
			{Optical fiber couplers-optimum solution for unequal cores},}\ }\href
	{\doibase 10.1109/50.4024} {\bibfield  {journal} {\bibinfo  {journal}
			{Journal of Lightwave Technology}\ }\textbf {\bibinfo {volume} {6}},\
		\bibinfo {pages} {463--474} (\bibinfo {year} {1988})}\BibitemShut {NoStop}%
	\bibitem [{\citenamefont {Huang}(1994)}]{Huang1994}%
	\BibitemOpen
	\bibfield  {author} {\bibinfo {author} {\bibfnamefont {W.-P.}\ \bibnamefont
			{Huang}},\ }\bibfield  {title} {\enquote {\bibinfo {title} {Coupled-mode
				theory for optical waveguides: an overview},}\ }\href {\doibase
		10.1364/JOSAA.11.000963} {\bibfield  {journal} {\bibinfo  {journal} {J. Opt.
				Soc. Am. A}\ }\textbf {\bibinfo {volume} {11}},\ \bibinfo {pages} {963--983}
		(\bibinfo {year} {1994})}\BibitemShut {NoStop}%
	\bibitem [{\citenamefont {Yamashita}\ \emph {et~al.}(1985)\citenamefont
		{Yamashita}, \citenamefont {Ozeki},\ and\ \citenamefont
		{Atsuki}}]{Yamashita1985}%
	\BibitemOpen
	\bibfield  {author} {\bibinfo {author} {\bibfnamefont {E.}~\bibnamefont
			{Yamashita}}, \bibinfo {author} {\bibfnamefont {S.}~\bibnamefont {Ozeki}}, \
		and\ \bibinfo {author} {\bibfnamefont {K.}~\bibnamefont {Atsuki}},\
	}\bibfield  {title} {\enquote {\bibinfo {title} {Modal analysis method for
				optical fibers with symmetrically distributed multiple cores},}\ }\href
	{\doibase 10.1109/JLT.1985.1074188} {\bibfield  {journal} {\bibinfo
			{journal} {Journal of Lightwave Technology}\ }\textbf {\bibinfo {volume}
			{3}},\ \bibinfo {pages} {341--346} (\bibinfo {year} {1985})}\BibitemShut
	{NoStop}%
	\bibitem [{\citenamefont {Kishi}\ \emph {et~al.}(1986)\citenamefont {Kishi},
		\citenamefont {Yamashita},\ and\ \citenamefont {Atsuki}}]{Kishi1986}%
	\BibitemOpen
	\bibfield  {author} {\bibinfo {author} {\bibfnamefont {Naoto}\ \bibnamefont
			{Kishi}}, \bibinfo {author} {\bibfnamefont {Eikichi}\ \bibnamefont
			{Yamashita}}, \ and\ \bibinfo {author} {\bibfnamefont {Kazuhiko}\
			\bibnamefont {Atsuki}},\ }\bibfield  {title} {\enquote {\bibinfo {title}
			{Modal and coupling field analysis of optical fibers with circularly
				distributed multiple cores and a central core},}\ }\href {\doibase
		10.1109/JLT.1986.1074869} {\bibfield  {journal} {\bibinfo  {journal} {Journal
				of Lightwave Technology}\ }\textbf {\bibinfo {volume} {4}},\ \bibinfo {pages}
		{991--996} (\bibinfo {year} {1986})}\BibitemShut {NoStop}%
	\bibitem [{\citenamefont {Kishi}\ and\ \citenamefont
		{Yamashita}(1988)}]{Kishi1988}%
	\BibitemOpen
	\bibfield  {author} {\bibinfo {author} {\bibfnamefont {Naoto}\ \bibnamefont
			{Kishi}}\ and\ \bibinfo {author} {\bibfnamefont {Eikichi}\ \bibnamefont
			{Yamashita}},\ }\bibfield  {title} {\enquote {\bibinfo {title} {A simple
				coupled-mode analysis method for multiple-core optical fiber and coupled
				dielectric waveguide structures},}\ }\href {\doibase 10.1109/22.17423}
	{\bibfield  {journal} {\bibinfo  {journal} {IEEE Transactions on Microwave
				Theory and Techniques}\ }\textbf {\bibinfo {volume} {36}},\ \bibinfo {pages}
		{1861--1868} (\bibinfo {year} {1988})}\BibitemShut {NoStop}%
	\bibitem [{\citenamefont {Schmidt-Hattenberger}\ \emph
		{et~al.}(1991)\citenamefont {Schmidt-Hattenberger}, \citenamefont
		{Trutschel}, \citenamefont {Muschall},\ and\ \citenamefont
		{Lederer}}]{Schmidt1991}%
	\BibitemOpen
	\bibfield  {author} {\bibinfo {author} {\bibfnamefont {C.}~\bibnamefont
			{Schmidt-Hattenberger}}, \bibinfo {author} {\bibfnamefont {U.}~\bibnamefont
			{Trutschel}}, \bibinfo {author} {\bibfnamefont {R.}~\bibnamefont {Muschall}},
		\ and\ \bibinfo {author} {\bibfnamefont {F.}~\bibnamefont {Lederer}},\
	}\bibfield  {title} {\enquote {\bibinfo {title} {Envelope description of an
				optical fibre array with circularly distributed multiple cores},}\ }\href
	{\doibase 10.1016/0030-4018(91)90361-G} {\bibfield  {journal} {\bibinfo
			{journal} {Opt. Commun.}\ }\textbf {\bibinfo {volume} {82}},\ \bibinfo
		{pages} {461} (\bibinfo {year} {1991})}\BibitemShut {NoStop}%
	\bibitem [{\citenamefont {Hudgings}\ \emph {et~al.}(2000)\citenamefont
		{Hudgings}, \citenamefont {Molter},\ and\ \citenamefont
		{Dutta}}]{Hudgings2000}%
	\BibitemOpen
	\bibfield  {author} {\bibinfo {author} {\bibfnamefont {Janice}\ \bibnamefont
			{Hudgings}}, \bibinfo {author} {\bibfnamefont {Lynne}\ \bibnamefont
			{Molter}}, \ and\ \bibinfo {author} {\bibfnamefont {Mitra}\ \bibnamefont
			{Dutta}},\ }\bibfield  {title} {\enquote {\bibinfo {title} {Design and
				modeling of passive optical switches and power dividers using non-planar
				coupled fiber arrays},}\ }\href {\doibase 10.1109/3.892564} {\bibfield
		{journal} {\bibinfo  {journal} {IEEE Journal of Quantum Electronics}\
		}\textbf {\bibinfo {volume} {36}},\ \bibinfo {pages} {1438--1444} (\bibinfo
		{year} {2000})}\BibitemShut {NoStop}%
	\bibitem [{\citenamefont {Rubenchik}\ \emph {et~al.}(2013)\citenamefont
		{Rubenchik}, \citenamefont {Tkachenko}, \citenamefont {Fedoruk},\ and\
		\citenamefont {Turitsyn}}]{Rubenchik2013}%
	\BibitemOpen
	\bibfield  {author} {\bibinfo {author} {\bibfnamefont {A~M}\ \bibnamefont
			{Rubenchik}}, \bibinfo {author} {\bibfnamefont {E~V}\ \bibnamefont
			{Tkachenko}}, \bibinfo {author} {\bibfnamefont {M~P}\ \bibnamefont
			{Fedoruk}}, \ and\ \bibinfo {author} {\bibfnamefont {S~K}\ \bibnamefont
			{Turitsyn}},\ }\bibfield  {title} {\enquote {\bibinfo {title}
			{Power-controlled phase-matching and instability of {CW} propagation in
				multicore optical fibers with a central core},}\ }\href {\doibase
		10.1364/OL.38.004232} {\bibfield  {journal} {\bibinfo  {journal} {Optics
				letters}\ }\textbf {\bibinfo {volume} {38}},\ \bibinfo {pages} {4232--5}
		(\bibinfo {year} {2013})}\BibitemShut {NoStop}%
	\bibitem [{\citenamefont {Hizanidis}\ \emph {et~al.}(2004)\citenamefont
		{Hizanidis}, \citenamefont {Droulias}, \citenamefont {Tsopelas},
		\citenamefont {Efremidis},\ and\ \citenamefont
		{Christodoulides}}]{Hizanidis2004}%
	\BibitemOpen
	\bibfield  {author} {\bibinfo {author} {\bibfnamefont {K.}~\bibnamefont
			{Hizanidis}}, \bibinfo {author} {\bibfnamefont {S.}~\bibnamefont {Droulias}},
		\bibinfo {author} {\bibfnamefont {I.}~\bibnamefont {Tsopelas}}, \bibinfo
		{author} {\bibfnamefont {N.~K.}\ \bibnamefont {Efremidis}}, \ and\ \bibinfo
		{author} {\bibfnamefont {D.~N.}\ \bibnamefont {Christodoulides}},\ }\bibfield
	{title} {\enquote {\bibinfo {title} {Centrally coupled circular array of
				optical waveguides: The existence of stable steady-state continuous waves and
				breathing modes},}\ }\href {\doibase 10.1238/Physica.Topical.107a00013}
	{\bibfield  {journal} {\bibinfo  {journal} {Physica Scripta}\ ,\ \bibinfo
			{pages} {13}} (\bibinfo {year} {2004})}\BibitemShut {NoStop}%
	\bibitem [{\citenamefont {Hizanidis}\ \emph {et~al.}(2006)\citenamefont
		{Hizanidis}, \citenamefont {Droulias}, \citenamefont {Tsopelas},
		\citenamefont {Efremidis},\ and\ \citenamefont
		{Christodoulides}}]{Hizanidis2006}%
	\BibitemOpen
	\bibfield  {author} {\bibinfo {author} {\bibfnamefont {K.}~\bibnamefont
			{Hizanidis}}, \bibinfo {author} {\bibfnamefont {S.}~\bibnamefont {Droulias}},
		\bibinfo {author} {\bibfnamefont {I.}~\bibnamefont {Tsopelas}}, \bibinfo
		{author} {\bibfnamefont {N.~K.}\ \bibnamefont {Efremidis}}, \ and\ \bibinfo
		{author} {\bibfnamefont {D.~N.}\ \bibnamefont {Christodoulides}},\ }\bibfield
	{title} {\enquote {\bibinfo {title} {Localized modes in a circular array of
				coupled nonlinear optical waveguides},}\ }\href {\doibase
		10.1142/S0218127406015647} {\bibfield  {journal} {\bibinfo  {journal} {Int.
				J. Bifur. Chaos}\ ,\ \bibinfo {pages} {1739}} (\bibinfo {year}
		{2006})}\BibitemShut {NoStop}%
	\bibitem [{\citenamefont {Radosavljevi}\ \emph {et~al.}(2015)\citenamefont
		{Radosavljevi}, \citenamefont {Danicic}, \citenamefont {Petrovic},
		\citenamefont {Maluckov},\ and\ \citenamefont
		{Hadzievski}}]{Radosavljevi2015}%
	\BibitemOpen
	\bibfield  {author} {\bibinfo {author} {\bibfnamefont {A}~\bibnamefont
			{Radosavljevi}}, \bibinfo {author} {\bibfnamefont {A}~\bibnamefont
			{Danicic}}, \bibinfo {author} {\bibfnamefont {J}~\bibnamefont {Petrovic}},
		\bibinfo {author} {\bibfnamefont {A}~\bibnamefont {Maluckov}}, \ and\
		\bibinfo {author} {\bibfnamefont {Lj.}\ \bibnamefont {Hadzievski}},\
	}\bibfield  {title} {\enquote {\bibinfo {title} {Coherent light propagation
				through multicore optical fibers with linearly coupled cores},}\ }\href
	{\doibase 10.1364/JOSAB.32.002520} {\bibfield  {journal} {\bibinfo  {journal}
			{Josa B}\ }\textbf {\bibinfo {volume} {32}},\ \bibinfo {pages} {2520--2527}
		(\bibinfo {year} {2015})}\BibitemShut {NoStop}%
	\bibitem [{\citenamefont {Mortimore}\ and\ \citenamefont
		{Arkwright}(1991)}]{Mortimore1991}%
	\BibitemOpen
	\bibfield  {author} {\bibinfo {author} {\bibfnamefont {D~B}\ \bibnamefont
			{Mortimore}}\ and\ \bibinfo {author} {\bibfnamefont {J~W}\ \bibnamefont
			{Arkwright}},\ }\bibfield  {title} {\enquote {\bibinfo {title} {Monolithic
				wavelength-flattened 1 x 7 single-mode fused fiber couplers: theory,
				fabrication, and analysis},}\ }\href {\doibase 10.1364/AO.30.000650}
	{\bibfield  {journal} {\bibinfo  {journal} {Applied Optics}\ }\textbf
		{\bibinfo {volume} {30}},\ \bibinfo {pages} {650--659} (\bibinfo {year}
		{1991})}\BibitemShut {NoStop}%
	\bibitem [{\citenamefont {Takenaga}\ \emph {et~al.}(2010)\citenamefont
		{Takenaga}, \citenamefont {Tanigawa}, \citenamefont {Guan}, \citenamefont
		{Matsuo}, \citenamefont {Saitoh},\ and\ \citenamefont
		{Koshiba}}]{Takenaga2010}%
	\BibitemOpen
	\bibfield  {author} {\bibinfo {author} {\bibfnamefont {K.}~\bibnamefont
			{Takenaga}}, \bibinfo {author} {\bibfnamefont {S.}~\bibnamefont {Tanigawa}},
		\bibinfo {author} {\bibfnamefont {N.}~\bibnamefont {Guan}}, \bibinfo {author}
		{\bibfnamefont {S.}~\bibnamefont {Matsuo}}, \bibinfo {author} {\bibfnamefont
			{K.}~\bibnamefont {Saitoh}}, \ and\ \bibinfo {author} {\bibfnamefont
			{M.}~\bibnamefont {Koshiba}},\ }\bibfield  {title} {\enquote {\bibinfo
			{title} {Reduction of crosstalk by quasi-homogeneous solid multi-core
				fiber},}\ }\href {\doibase 10.1364/OFC.2010.OWK7} {\bibfield  {journal}
		{\bibinfo  {journal} {Optical Fiber Communication (OFC), collocated National
				Fiber Optic Engineers Conference, 2010 Conference on (OFC/NFOEC)}\ ,\
			\bibinfo {pages} {35--37}} (\bibinfo {year} {2010})}\BibitemShut {NoStop}%
	\bibitem [{\citenamefont {Koshiba}\ \emph {et~al.}(2011)\citenamefont
		{Koshiba}, \citenamefont {Saitoh}, \citenamefont {Takenaga},\ and\
		\citenamefont {Matsuo}}]{Koshiba2011}%
	\BibitemOpen
	\bibfield  {author} {\bibinfo {author} {\bibfnamefont {Masanori}\
			\bibnamefont {Koshiba}}, \bibinfo {author} {\bibfnamefont {Kunimasa}\
			\bibnamefont {Saitoh}}, \bibinfo {author} {\bibfnamefont {Katsuhiro}\
			\bibnamefont {Takenaga}}, \ and\ \bibinfo {author} {\bibfnamefont
			{Shoichiro}\ \bibnamefont {Matsuo}},\ }\bibfield  {title} {\enquote {\bibinfo
			{title} {Multi-core fiber design and analysis: coupled-mode theory and
				coupled-power theory},}\ }\href {\doibase 10.1364/OE.19.00B102} {\bibfield
		{journal} {\bibinfo  {journal} {Optics Express}\ }\textbf {\bibinfo {volume}
			{19}},\ \bibinfo {pages} {B102} (\bibinfo {year} {2011})}\BibitemShut
	{NoStop}%
	\bibitem [{\citenamefont {Chan}\ \emph {et~al.}(2012)\citenamefont {Chan},
		\citenamefont {Lau},\ and\ \citenamefont {Tam}}]{Chan2012}%
	\BibitemOpen
	\bibfield  {author} {\bibinfo {author} {\bibfnamefont {Florence Y.~M.}\
			\bibnamefont {Chan}}, \bibinfo {author} {\bibfnamefont {Alan Pak~Tao}\
			\bibnamefont {Lau}}, \ and\ \bibinfo {author} {\bibfnamefont {Hwa-Yaw}\
			\bibnamefont {Tam}},\ }\bibfield  {title} {\enquote {\bibinfo {title} {Mode
				coupling dynamics and communication strategies for multi-core fiber
				systems},}\ }\href {\doibase 10.1364/OE.20.004548} {\bibfield  {journal}
		{\bibinfo  {journal} {Optics Express}\ }\textbf {\bibinfo {volume} {20}},\
		\bibinfo {pages} {4548} (\bibinfo {year} {2012})}\BibitemShut {NoStop}%
	\bibitem [{\citenamefont {Chekhovskoy}\ \emph {et~al.}(2017)\citenamefont
		{Chekhovskoy}, \citenamefont {Sorokina}, \citenamefont {Rubenchik},\ and\
		\citenamefont {Fedoruk}}]{Chekhovskoy2017}%
	\BibitemOpen
	\bibfield  {author} {\bibinfo {author} {\bibfnamefont {I.S.}\ \bibnamefont
			{Chekhovskoy}}, \bibinfo {author} {\bibfnamefont {M.A.}\ \bibnamefont
			{Sorokina}}, \bibinfo {author} {\bibfnamefont {A.M.}\ \bibnamefont
			{Rubenchik}}, \ and\ \bibinfo {author} {\bibfnamefont {M.P.}\ \bibnamefont
			{Fedoruk}},\ }\bibfield  {title} {\enquote {\bibinfo {title} {Spatiotemporal
				multiplexing based on hexagonal multicore optical fibres},}\ }\href {\doibase
		10.1070/QEL16536} {\bibfield  {journal} {\bibinfo  {journal} {Quantum
				Electronics}\ }\textbf {\bibinfo {volume} {47}} (\bibinfo {year} {2017}),\
		10.1070/QEL16536}\BibitemShut {NoStop}%
	\bibitem [{\citenamefont {Hossain}\ and\ \citenamefont
		{Majumder}(2017)}]{Hossain2017}%
	\BibitemOpen
	\bibfield  {author} {\bibinfo {author} {\bibfnamefont {M.~A.}\ \bibnamefont
			{Hossain}}\ and\ \bibinfo {author} {\bibfnamefont {S.~P.}\ \bibnamefont
			{Majumder}},\ }\bibfield  {title} {\enquote {\bibinfo {title} {Performance
				analysis of crosstalk due to inter-core coupling in a heterogeneous
				multi-core optical fiber communication system},}\ }in\ \href {\doibase
		10.1109/ICTP.2017.8285912} {\emph {\bibinfo {booktitle} {2017 IEEE
				International Conference on Telecommunications and Photonics (ICTP)}}}\
	(\bibinfo  {publisher} {IEEE},\ \bibinfo {year} {2017})\ pp.\ \bibinfo
	{pages} {142--146}\BibitemShut {NoStop}%
	\bibitem [{\citenamefont {Vance}(1996)}]{Vance1996}%
	\BibitemOpen
	\bibfield  {author} {\bibinfo {author} {\bibfnamefont {R.}~\bibnamefont
			{Vance}},\ }\bibfield  {title} {\enquote {\bibinfo {title} {Matrix {Lie}
				group-theoretic design of coupled linear optical waveguide devices},}\ }\href
	{\doibase 10.1137/S0036139994269923} {\bibfield  {journal} {\bibinfo
			{journal} {SIAM Journal on Applied Mathematics}\ }\textbf {\bibinfo {volume}
			{56}},\ \bibinfo {pages} {765} (\bibinfo {year} {1996})}\BibitemShut
	{NoStop}%
	\bibitem [{\citenamefont {Rodr\'{i}guez-Lara}\ and\ \citenamefont
		{Guerrero}(2015)}]{RodriguezLara2015}%
	\BibitemOpen
	\bibfield  {author} {\bibinfo {author} {\bibfnamefont {B.~M.}\ \bibnamefont
			{Rodr\'{i}guez-Lara}}\ and\ \bibinfo {author} {\bibfnamefont
			{J.}~\bibnamefont {Guerrero}},\ }\bibfield  {title} {\enquote {\bibinfo
			{title} {Optical finite representation of the lorentz group},}\ }\href
	{\doibase 10.1364/OL.40.005682} {\bibfield  {journal} {\bibinfo  {journal}
			{Opt. Lett.}\ }\textbf {\bibinfo {volume} {40}},\ \bibinfo {pages} {5682}
		(\bibinfo {year} {2015})},\ \Eprint {http://arxiv.org/abs/1508.05419}
	{arXiv:1508.05419 [physics.optics]} \BibitemShut {NoStop}%
	\bibitem [{\citenamefont {Vergara}\ and\ \citenamefont
		{Rodr\'{i}guez-Lara}(2015)}]{Vergara2015}%
	\BibitemOpen
	\bibfield  {author} {\bibinfo {author} {\bibfnamefont {L.~Villanueva}\
			\bibnamefont {Vergara}}\ and\ \bibinfo {author} {\bibfnamefont {B.~M.}\
			\bibnamefont {Rodr\'{i}guez-Lara}},\ }\bibfield  {title} {\enquote {\bibinfo
			{title} {Gilmore-perelomov symmetry based approach to photonic lattices},}\
	}\href {\doibase 10.1364/OE.23.022836} {\bibfield  {journal} {\bibinfo
			{journal} {Opt. Express}\ }\textbf {\bibinfo {volume} {23}},\ \bibinfo
		{pages} {22836} (\bibinfo {year} {2015})},\ \Eprint
	{http://arxiv.org/abs/1506.02062} {arXiv:1506.02062 [physics.optics]}
	\BibitemShut {NoStop}%
	\bibitem [{\citenamefont {Huerta~Morales}\ \emph {et~al.}(2016)\citenamefont
		{Huerta~Morales}, \citenamefont {Guerrero}, \citenamefont {López-Aguayo},\
		and\ \citenamefont {Rodríguez-Lara}}]{Huerta2016}%
	\BibitemOpen
	\bibfield  {author} {\bibinfo {author} {\bibfnamefont {J.~D.}\ \bibnamefont
			{Huerta~Morales}}, \bibinfo {author} {\bibfnamefont {J.}~\bibnamefont
			{Guerrero}}, \bibinfo {author} {\bibfnamefont {S.}~\bibnamefont
			{López-Aguayo}}, \ and\ \bibinfo {author} {\bibfnamefont {B.~M.}\
			\bibnamefont {Rodríguez-Lara}},\ }\bibfield  {title} {\enquote {\bibinfo
			{title} {Revisiting the optical {PT}-symmetric dimer},}\ }\href {\doibase
		10.3390/sym8090083} {\bibfield  {journal} {\bibinfo  {journal} {Symmetry}\
		}\textbf {\bibinfo {volume} {8}} (\bibinfo {year} {2016}),\
		10.3390/sym8090083},\ \Eprint {http://arxiv.org/abs/1607.02782}
	{arXiv:1607.02782 [physics.optics]} \BibitemShut {NoStop}%
	\bibitem [{\citenamefont {{Nodal Stevens}}\ \emph {et~al.}(2018)\citenamefont
		{{Nodal Stevens}}, \citenamefont {{Jaramillo \'{A}vila}},\ and\ \citenamefont
		{Rodr\'{i}guez-Lara}}]{NodalStevens2018}%
	\BibitemOpen
	\bibfield  {author} {\bibinfo {author} {\bibfnamefont {D.~J.}\ \bibnamefont
			{{Nodal Stevens}}}, \bibinfo {author} {\bibfnamefont {Benjam\'{i}n}\
			\bibnamefont {{Jaramillo \'{A}vila}}}, \ and\ \bibinfo {author}
		{\bibfnamefont {B.~M.}\ \bibnamefont {Rodr\'{i}guez-Lara}},\ }\bibfield
	{title} {\enquote {\bibinfo {title} {Necklaces of {PT}-symmetric dimers},}\
	}\href {\doibase 10.1364/PRJ.6.000A31} {\bibfield  {journal} {\bibinfo
			{journal} {Photon. Res.}\ }\textbf {\bibinfo {volume} {6}},\ \bibinfo {pages}
		{A31} (\bibinfo {year} {2018})},\ \Eprint {http://arxiv.org/abs/1709.00498}
	{arXiv:1709.00498 [physics.optics]} \BibitemShut {NoStop}%
	\bibitem [{\citenamefont {Rodr\'{i}guez-Lara}\ \emph
		{et~al.}(2018)\citenamefont {Rodr\'{i}guez-Lara}, \citenamefont
		{El-Ganainy},\ and\ \citenamefont {Guerrero}}]{RodriguezLara2018}%
	\BibitemOpen
	\bibfield  {author} {\bibinfo {author} {\bibfnamefont {B.~M.}\ \bibnamefont
			{Rodr\'{i}guez-Lara}}, \bibinfo {author} {\bibfnamefont {Ramy}\ \bibnamefont
			{El-Ganainy}}, \ and\ \bibinfo {author} {\bibfnamefont {Julio}\ \bibnamefont
			{Guerrero}},\ }\bibfield  {title} {\enquote {\bibinfo {title} {Symmetry in
				optics and photonics: a group theory approach},}\ }\href {\doibase
		10.1016/j.scib.2017.12.020} {\bibfield  {journal} {\bibinfo  {journal}
			{Science Bulletin}\ }\textbf {\bibinfo {volume} {63}},\ \bibinfo {pages}
		{244} (\bibinfo {year} {2018})},\ \Eprint {http://arxiv.org/abs/1803.00121}
	{arXiv:1803.00121 [quant-ph]} \BibitemShut {NoStop}%
	\bibitem [{\citenamefont {Choi}\ \emph {et~al.}(2017)\citenamefont {Choi},
		\citenamefont {Park}, \citenamefont {Elliott},\ and\ \citenamefont
		{Oh}}]{Choi2017}%
	\BibitemOpen
	\bibfield  {author} {\bibinfo {author} {\bibfnamefont {H.}~\bibnamefont
			{Choi}}, \bibinfo {author} {\bibfnamefont {M.}~\bibnamefont {Park}}, \bibinfo
		{author} {\bibfnamefont {D.~S.}\ \bibnamefont {Elliott}}, \ and\ \bibinfo
		{author} {\bibfnamefont {K.}~\bibnamefont {Oh}},\ }\bibfield  {title}
	{\enquote {\bibinfo {title} {Optomechanical measurement of the abraham force
				in an adiabatic liquid-core optical-fiber waveguide},}\ }\href {\doibase
		10.1103/PhysRevA.95.053817} {\bibfield  {journal} {\bibinfo  {journal} {Phys.
				Rev. A}\ }\textbf {\bibinfo {volume} {95}},\ \bibinfo {pages} {053817}
		(\bibinfo {year} {2017})},\ \Eprint {http://arxiv.org/abs/1601.05225}
	{arXiv:1601.05225 [physics.optics]} \BibitemShut {NoStop}%
\end{thebibliography}

\end{document}